\documentclass[letterpaper]{article} 
\usepackage{aaai24}  
\usepackage{times}  
\usepackage{helvet}  
\usepackage{courier}  
\usepackage[hyphens]{url}  
\usepackage{graphicx} 
\usepackage{natbib}  
\usepackage{caption} 
\usepackage{algorithm}
\usepackage{algorithmic}

\usepackage{newfloat}
\usepackage{listings}
\DeclareCaptionStyle{ruled}{labelfont=normalfont,labelsep=colon,strut=off} 
\floatstyle{ruled}
\newfloat{listing}{tb}{lst}{}
\floatname{listing}{Listing}
\title{Misrepresented Technological Solutions in Imagined Futures: The Origins and Dangers of AI Hype in the Research Community}
\author{
Savannah Thais
}
\affiliations{
Columbia University Data Science Institute\\

    New York, New York 11221 USA\\
    st3565@columbia.edu\\
}

\usepackage{bibentry}

\begin{document}

\maketitle

\begin{abstract}
Technology does not exist in a vacuum; technological development, media representation, public perception, and governmental regulation cyclically influence each other to produce the collective understanding of a technology's capabilities, utilities, and risks. When these capabilities are overestimated, there is an enhanced risk of subjecting the public to dangerous or harmful technology, artificially restricting research and development directions, and enabling misguided or detrimental policy. The dangers of technological hype are particularly relevant in the rapidly evolving space of AI. Centering the research community as a key player in the development and proliferation of hype, we examine the origins and risks of AI hype to the research community and society more broadly and propose a set of measures that researchers, regulators, and the public can take to mitigate these risks and reduce the prevalence of unfounded claims about the technology. \end{abstract}

\section{Introduction}
Artificial Intelligence (AI) has become a ubiquitous topic in the modern world as it has captivated the popular imagination, spurred rapid technological development, and captured a significant share of the investment market. However, despite being presumably a scientific field, many of the descriptions and promises of AI systems are not empirically proven or conceptually well grounded. 

The media, and even some researchers themselves, unfoundedly ascribe language understanding, general reasoning ability, or even sentience to AI systems, and claim that we are close to achieving artificial general intelligence (AGI) \cite{expected_agi}. 
Through various intentional or accidental mechanisms including failures in testing or system design \cite{leakage}, poorly constructed or characterized data sets \cite{everything_benchmark, fallacy}, or outright deception \cite{fallacy}, researchers and developers can create AI systems that fail to live up to their performance and reliability promises.

This culture of hype has real world consequences. Inaccurate or misrepresented AI systems can cause harm to the public and exacerbate societal biases at an unprecedented scale \cite{fallacy,wang_against}. Unscientific and misleading claims about AI also negatively impact the research and development ecosystem itself by incentivizing certain research directions over others and affecting how broader society views the validity and utility of the field. 

Here, we define AI hype as any non empirically or rigorously theoretically supported performance claims, capability narratives, or system descriptions. We consider empirically supported claims to mean performances or capabilities that demonstrated through properly designed, reproducible scientific research, that generalize outside of the initial experimental context, and are precisely characterized in their descriptive language (i.e. not saying a model exhibits language understanding when what is meant is that it achieved high accuracy on a multiple choice benchmark data set like MMLU \cite{mmlu}), while we consider theoretically supported claims to be provably derived directly from statistical or mathematical theory. 

Centering the research and development community as a key responsible player in the development and proliferation of hype, this paper will explore origins of AI hype and the dangers that hype poses to both society and the research ecosystem. We will then present actions that the research community can take to address these concerns and ways that those efforts can be scaffolded by political action and community work.

\section{Current Landscape of AI Rhetoric}
Recent advancements in AI, though undoubtedly impressive, have been surrounded by unprecedented levels of hype. Particular attention has been paid recently to large language models (LLMs). Some claims like scientific reasoning and understanding of the short-lived model Galactica \cite{galactica} or the ascribed sentience of more general LLMs like LaDMa \cite{ladma} received relatively little uptake or have been quickly debunked \cite{galactica_mit,llm_sentience}.

More pernicious and common are claims of general or approximate language understanding\cite{nlu_climbing} and reasoning abilities \cite{sparks_agi}. This belief enables additional sociological claims that LLMs will soon replace human writers in a variety of domains \cite{writers_artists_obsolete}, precipitate a breakdown of traditional education \cite{chatgpt_college_essay}, and transform search and information retrieval \cite{situating_search}, which, while entirely possible due to financial power structures, are largely ungrounded in scientifically proven abilities of these models (indeed, we have already seen the threat of AI replacement play a central role in the Hollywood writers strike \cite{writers_strike}). While LLMs have demonstrated exceptional performance on certain purported reasoning related benchmark data sets, such as MMLU, they perfrom substantially worse than humans on others, such as ConceptARC \cite{conceptarc} and PlanBench \cite{valmeekam2023planbench}. As discussed further in Section \ref{sec:randd}, these apparent discrepancies in ability are in part attributable to the design and conceptualization of benchmark tasks themselves.

Furthermore, research in NLP has demonstrated that despite these LLMs performing well on some established benchmark tasks, this performance can be attributed to learning to manipulate established linguistic forms \cite{nlu_climbing} or to leverage artifacts from the training data sets \cite{nlp_adversarial1,nlp_adversarial2}, and in fact this performance often does not generalize. For example, Niven and Kao demonstrated that BERT's performance on the Argument Reasoning Comprehension Task is accounted for by exploitation of spurious statistical cues in the data set \cite{nlp_adversarial1}.

Moreover, these claims obfuscate known limitations and issues with the technology. Notably, they are prone to hallucination and can generate coherent and convincing yet false text, including factually incorrect historical or scientific facts, causally inaccurate reasoning, or fabricated sources and references \cite{chatgpt_abstract,medical_voiceassistant,chatgpt_falsehoods}. Additionally, they often demonstrate bias against certain groups, owing in large part to their training corpus of text scraped from the internet \cite{stochastic_parrots,gpt3_official,chatbot_safetyrecs,languagemodel_muslim_bias}. Unfortunately, some initial exploration demonstrates that these issues persist even in models that are nominally safeguarded, such as ChatGPT \cite{chatgpt_bias}. These substantial limitations and the risks they pose, as described in Section 4, currently render them largely non-robust and untrustworthy, and perhaps of limited practical utility. 

Similar hype and controversy surround other types of generative AI, namely text-to-image and text-to-video generators like DALL-E 2, Stable Diffusion, Make-A-Video, Imagen Video, and others. These models have been described as demonstrating visual reasoning skills or human-like creativity \cite{generativeai_mittech}. As with LLMs, news outlets, researchers, and artists themselves have questioned if, or even predicted that, this technology will soon replace (or at least substantially transform) a wide range of creative careers \cite{generativeai_mittech, writers_artists_obsolete}.

While the capabilities of these generative AI models are undoubtedly impressive, many of the claims surrounding this technology are, again, not empirically grounded. In particular, recent research has demonstrated that these models are unable to correctly interpret prompts containing specific spatial relation or object multiplicity information \cite{diffusion_reasoning1,diffusion_reasoning2,diffusion_reasoning_bias}. Furthermore, they are not robust to prompt variation, and small adjustments or re-orderings of the prompt can result in substantially different outputs \cite{diffusion_prompt_engineering, diffusion_reasoning2}. Additionally, they are susceptible to the same issues of bias present in LLMs; they demonstrate gender and skin-tone biases when given neutral text prompts \cite{diffusion_reasoning_bias} as well as context bias when provided with gender or race specified input \cite{lensa_nude}.

Although there is substantial hype around recent AI systems, this phenomena is by no means new. Autonomous driving systems are perhaps the most over-hyped AI technology. Various executives and researchers have claimed for years that fully autonomous or self-driving vehicles are on the verge of entering the market; for example, in 2017 Jensen Huang, the CEO of Nvidia, proclaimed that ``we can realize this vision [of self-driving cars] right now'', clarifying that ``AI is the solution to self-driving'' \cite{ml_selfdriving}. Two years previously, Elon Musk, the CEO of Tesla, promised that self-driving is ``almost...a solved problem''; Musk has continued to tout Tesla's self-driving capabilities since, despite having to repeatedly walk back, temper, or qualify these claims \cite{ml_selfdriving}. Regardless of the continued hype and large-scale financial investments, only a handful of prototype vehicles have been tested on public roads, and safe and reliable autonomous vehicles remain beyond the immediate horizon of technological capabilities \cite{selfdriving_faroff}. In fact, backlash to the sustained hype around this technology and its lack of realization has led some to conclude that ``self-driving cars are going nowhere'' \cite{selfdriving_nowhere}.

AI hype is not limited to over-estimations of the ability of AI systems to perform on general reasoning tasks. Many focused or narrow applications of AI are subject to misguided or misleading narratives. AI systems have been purported to reliably perform on a diverse array of tasks, ranging from predicting someone's appearance, political affiliation, criminality, and more from their voice or physical attributes \cite{speech2face,ai_physigonomy} to triaging patients \cite{hospitalbed_bias}, allocating welfare benefits \cite{welfare_ai}, and selecting employees for layoffs \cite{litigating_algorithms}. Unfortunately, the performance claims surrounding many of these systems are not empirically proven or are not replicated when the systems are deployed in the real world. 

As described in the following section, some of these unsubstantiated claims can be attributed to engineering or design mistakes that were not noticed during development. However, a substantial portion of the inaccurate rhetoric surrounding this type of AI is due to attempts to apply AI to what Raji et al term ``conceptually impossible tasks'' where no causal connection exists between the observable data and the proposed task \cite{fallacy}, despite developers' or marketers' claims that AI can learn information that simply is not perceived by the human brain.

The culture of hype surrounding different types of AI systems, as well as factors described in the next section, contribute to perhaps the most ubiquitous non-scientifically grounded belief about AI: the expectation of near future AGI. Although there is no agreed upon scientific definition of AGI \cite{levels_agi} and AI researchers hold extremely varied opinions about the near-term likelihood of AGI \cite{researcher_survey} (in fact, many have stated that AGI is far from being realized \cite{beyond_longterm, its_goingto_killus}), this rhetoric has captured substantial public and media attention, garnered significant financial investment, and enabled the proliferation of AGI focused start-ups.

\section{Origins of AI Hype}
In order to fully characterize the dangers associated with AI hype and identify potential mitigating strategies, we must first understand the factors that allow hype to develop and spread. While many analyses of technological hype focus on popular media representations or financial incentives as origins of hype, here we will focus largely on issues inherent to the current AI research and development (R\&D) ecosystem and explore how these concerns enable and enhance broader hype cycles.

AI researchers and developers exercise nearly complete control over the design, and thereby applicability and reliability, of a given AI system. Nonetheless, they typically perceive both the underlying data and eventual impacts of these systems as outside of their control \cite{Slota2020}. Along with the issues described in the following section, this fosters a culture of distance between R\&D and the societal context of technology. This can lead to researchers and developers prioritizing certain technical considerations like efficiency over more human-centered concerns like preventing harm \cite{normative_uncertainty}. This perceived distance becomes particularly problematic in the current fast-paced technological R\&D ecosytem that tends to prioritize the race to innovation and the capture of market share.     

By focusing primarily on the R\&D ecosystem, we hope to re-center researchers and developers as critical actors in the construction and propagation of misleading or dangerous rhetoric around AI and shift some of the responsibility for counteracting these narratives back to these groups.

\subsection{Research and Development Environments}\label{sec:randd}
AI hype originates in part from the culture of R\&D environments and how researchers and developers mathematically and scientifically characterize current AI technology. Many of the issues outlined below are fundamentally intertwined; responsible and honest R\&D must reckon with these considerations collectively.

A primary origin of over-hyped AI capabilities is the fact that many AI systems are developed in sterile R\&D environments and then deployed in more complex real-world settings without appropriate testing or oversight. In some cases, a mismatch between performance in R\&D and deployment settings is attributable to sampling bias in the test set distribution \cite{leakage}. Choices made during data collection and pre-processing can render a data set non-representative of the population the AI system is designed to be used on; for example, a disease prediction tool trained on medical images may not generalize from one hospital environment to another \cite{medicalimages_hospitaltransfer} or may not be robust to small perturbations caused by malfunctions of the imaging device \cite{medicalimages_variability}.

In other cases, performance disparities are attributable to a combination of data set and evaluation metric design choices. Namely, discrepancies between evaluations in an R\&D settings and a meaningful characterization of how the AI system will actually perform in the real world \cite{everything_benchmark}. A lack of so-called `construct validity' can lead to performance issues by allowing researchers and developers to neglect to include distribution of societal harms and benefits in their evaluation metrics \cite{fallacy} and by focusing R\&D efforts on improving performance on artificially constructed benchmark data sets that do not meaningfully relate to the general reasoning tasks they purport to measure \cite{everything_benchmark}.

Designing effective and robust benchmark data sets for AI models is a challenging task, and mischaracterizing what a benchmark is designed to do is a significant contributor to AI hype. For example, ImageNet is widely accepted as a benchmark for visual object recognition, yet it seemingly arbitrarily relies on the WorldNet ontology to derive class labels, and thus is highly limited in the classes it actually represents. It has therefore been critiqued as actually representing a very specifc, rather than general, vision benchmark \cite{everything_benchmark}.

Some R\&D environments and applications of AI also eschew critical scientific and software development practices. For example, as demonstrated by surveys of AI researchers, the predictions made by AI systems are often taken at face value without a meaningful understanding of the associated uncertainty \cite{Slota2020, transparency_in_computing}. In other scientific fields, the uncertainty of a measurement or model is an essential component of understanding its significance and utility. Neglecting to provide accounts of the uncertainty associated with AI systems allows developers to gloss over potential concerns with the underlying training data, possible mismatch between the predicted labels and the real world decision being modeled, and many other critical design issues \cite{fallacy,leakage,wang_against}. Additionally, studies have found that many AI developers often disregard traditional software development practices like testing and quality assurance \cite{ml_dev_practices} and that ML and AI development occurs substantial `technical debt' in terms of documentation, reproducibility, and reliability \cite{technical_debt}. These issues coalesce to create the present so-called `reproducibility crisis' in AI and ML R\&D. 

There are also issues inherent to the culture of technology R\&D including how technologists view other fields and how we cognitively and linguistically frame the problems that we work on. Many AI systems seek to make predictions about complex social behaviors and future outcomes and many researchers and developers hold a belief that such predictions are possible with currently available data. However, as Campolo and Crawford describe, this belief requires an ``epistemological flattening'' of complex societal contexts into acute signals that neglect the probabilistic worldviews traditionally employed in social science research \cite{enchanted_determinism}. As demonstrated in a study by Sambasivan and Veeraraghavan, AI researchers and developers also tend to neglect the importance of domain expertise in understanding the appropriate development and limitations of AI systems and frequently reduce researchers and practitioners in other fields to data collectors rather than equal collaborators \cite{deskilling}. This domain hubris can result in a host of technical issues including feature illegitimacy \cite{feature_legitimacy}, lack of appropriate data segregation \cite{data_segregation}, and insufficient pre-deployment testing \cite{medical_validity_testing}.

Additionally, despite some advances in explainable and interpretable AI, there remains a profound mismatch between the high-dimensionality of many AI systems and the interpretation capacity of human developers and users \cite{Burrell2016-gy}. There is a profound disconnect between technological progress and a fundamental scientific understanding of how AI systems function \cite{enchanted_determinism}. This leads researchers and developers to describe their work in less precise or exact terminology and creates a perceived lack of control over the technological narrative \cite{Slota2020}. As discussed further in Section 3.3, the language used by researchers and developers to describe their work can substantially contribute to AI hype.

\subsection{Financial Incentives}
The R\&D ecosystem and its outputs cannot be divorced from the financial mechanisms that support it. Technological R\&D is often viewed as a primary driver of economic prosperity and national security \cite{brookings_tech_economy} and thus attracts substantial public and private investment. AI R\&D in particular currently dominates global scientific output \cite{unesco_science_report} and continues to attract increased funding; the private investment in AI was 18 times higher in 2022 than in 2013, totaling around \$93.5 billion \cite{hai_report}. Unlike some other areas of scientific R\&D, the AI ecosystem is dominated by this private money; in fact, many `top' academic researchers receive substantial financial support from private tech companies \cite{bigtech_in_academia}. This funding landscape can incentivize hype and skew publishing norms \cite{aiethics_global_community}.

In particular, the relative abundance of funding but expected short timelines for innovation and return on investment lead to what Rayner calls the `novelty trap'. Researchers and developers are incentivized to exaggerate the speed and inventiveness of their work while de-emphasizing potential safety and reliability concerns \cite{novelty_trap}. Insidiously, this means that there is often substantial profit to be earned by marketing technological abilities that may not exist \cite{boyd2018}. For example, Chinese technology company Baidu was discovered to have cheated on the popular benchmark data set ImageNet in an effort to include exaggerated performance claims in marketing materials \cite{baidu_cheating}; similarly, Amazon misrepresented the ways in which their facial recognition software, Rekognition, could be used by law enforcement in sales and promotional material \cite{aclu_rekognition}. This type of behavior dangerously shifts the responsibility for critically evaluating system performance claims to consumers, regulators, and external researchers, who are often less equipped to undertake this work or are not given necessary access to the systems in question. For example, the landmark Gender Shades study, conducted by external researchers Buolamwini and Gebru uncovered gender and racial bias in commercial facial recognition technology \cite{gendershades} and was essential for incentivizing the involved tech companies to make their systems safer \cite{gendershades_audit}.

\subsection{Media and Society}
Additionally, R\&D and the surrounding financial landscape are impacted by public and media discourse and representations of AI. AI has been part of collective human fantasy and science fiction narratives long before technological advances made any of these imagined capabilities remotely possible. These depictions and the resulting cultural norms formed around AI influence not only how the public and reporters perceive this technology, but also how researchers and developers imagine the capabilities and limitations of the very technology they are creating.

These normative representations, as well as a real lack of deep mathematical understanding of how many modern AI systems (particularly deep learning systems) function \cite{expressivepower_nns,expressivepower_nns2}, impact how the individuals closest to AI describe their work. Both developers \cite{enchanted_determinism} and critics \cite{critics_hype} of technology employ magical, anthropomorphic, or superhuman language to describe AI systems. For example, in the official paper for AlphaZero, a Go-playing AI system, researchers described the technology as ``superhuman''  and operating ``without human knowledge'' despite the fact that the entire system was, in fact, developed by humans with an understanding of the mechanisms and goals of the game \cite{alphazero}. Similarly, Demis Hassabis, the CEO of DeepMind, the company behind AlphaZero, likened the system to a chess-playing alien \cite{alphazero_alien}.

It is unsurprising, then, that journalists reporting on AI face difficulties in accurately discussing the technology. This is further complicated by the reality that journalists may lack the technological and scientific foundations to fully understand and characterize AI even in cases when it is precisely and truthfully described by researchers and developers \cite{aijournalism_jones}. Some researchers and developers have described concerns that outcomes and nuances of AI research are distorted as they leave the immediate R\&D ecosystem and cite concern over a lack of scientific precision in media coverage \cite{Slota2020}.

Together, these factors can foster a cycle of misrepresentation where R\&D environments, funding mechanisms, and social and media representations influence each other to create a collective understanding of AI that may be distant from the scientific reality.

\subsection{Technologies as Imagined Futures}
When interrogating AI hype, it is crucial to emphasize that rhetoric around technology, regardless of the source, is discursive world building. Technology not only represents a solution to a problem, but through its development and presentation asserts what problems exist and should be considered. Stilgoe discusses this phenomena in the context of autonomous vehicles: in media coverage and public policy, vehicular accidents have traditionally been blamed on driver error, and by taking up this rhetoric, autonomous vehicle companies are able to present self-driving cars and AI as a solution to the ``driver problem''. By doing so, they construct an imagined future that entrenches consumer reliance on individual vehicles and detracts financial and intellectual resources from other potential solutions like improved public transit \cite{politics_autonomousvehicles}.

Shah and Bender similarly position claims that LLMs will replace search systems. These claims, often perpetuated by the developers of the LLMs in question, neglect to consider the societal context and purpose of search. In many cases, it is more useful and desirable for users to explore and make sense of information on their own, rather than be provided with a singular answer as is common with LLMs \cite{situating_search}. By promoting a certain narrative about the abilities and utility of technology, we necessarily impact societal behavior and views about technology.

\section{Dangers of Misrepresentation}
The pervasiveness of AI hype can prevent researchers and developers, policy makers, journalists, and the public from proactively interrogating these imagined futures. Hype can foster the impression that AI is universally good or bad and can preclude the nuance necessary to use this technology effectively \cite{Slota2020}. Similarly, claims of novelty and innovation can hinder social learning and proper contextualization of AI and create a perceived distance between the technology and its risks \cite{novelty_trap,ml_selfdriving}. 

Campolo and Crawford describe AI hype as `enchanted determinism', allowing AI systems to be seen as powerfully predictive and deterministic, yet also as outside the realm of full human understanding and ability \cite{enchanted_determinism}. The unique qualities of this type of hype make it particularly persuasive to a broad range of stakeholders. Its power and ubiquity pose substantial risks not only to society at large but also, critically, to the very R\&D ecosystem in which AI technology is constructed.

\subsection{Unscientific, Inaccurate, and Dangerous Sociotechnical Systems}
When AI capabilities are mischaracterized, there is heightened risk of causing harm to individuals who rely on or are subjected to the predictions of these systems. There are already numerous examples of tangible financial, emotional, and corporeal harm affected by misrepresented or misused AI systems. For example, voice assistants such as Alexa, Siri, and Google Assistant, have been portraye having the potential to serve as portals for health information \cite{va_health}. However, a study found that a when given medical related queries, a substantial portion of the information these voice assistants returned could result in user harm or death \cite{medical_voiceassistant}. In another case, an external audit of a widely deployed sepsis prediction tool developed by the healthcare software company Epic found that despite performance claims to the contrary, the tool failed to identify a majority of sepsis cases and generated a considerable number of false alerts, creating extraordinary risk for treatment mistakes \cite{sepsis}. In yet another example, Allegheny county in Pennsylvania adopted a screening tool designed to identify children at risk of mistreatment and used it to select families for investigations that could involve removing the child from their families. A full two years after the tool was deployed it was discovered that reported performance measurements were over estimated and that the tool exhibited substantially higher error rates on Black families compared to White families \cite{afst}. 

The widespread utilization of this type of unreliable predictive AI is directly enabled by AI hype. The specific type of hype surrounding AI, as described in Section 3.3, is used to justify a movement away from explainable decision systems and reduced focus on clear causal mechanisms in high-stakes AI enabled predictions \cite{enchanted_determinism}. The deployment of uninterpretable AI denies the people subjected to these systems opportunities for recourse. 

Furthermore, AI hype fosters a culture of R\&D and AI system utilization that values ostensible predictive accuracy over deeper contextual or scientific knowledge of the underlying mechanisms or potential impacts of those predictions \cite{enchanted_determinism}. This allows researchers and developers to ignore validity issues intrinsic to the structure of predictive optimization tasks which Wang et al demonstrate carry inherent risks of harm to individuals \cite{wang_against}. 

The way these systems are represented and marketed often results in a displacement of responsibility for negative outcomes onto individual users or subjects who, in practice, lack control over the outcomes \cite{Slota2020,crumplezones}. For example, despite emphasizing the predictive power, accuracy, and automation capabilities of their systems, technology companies often attempt to distance themselves from harmful impacts of their technology by claiming that the systems were never intended to be used as decision tools and require extensive human oversight to be used safely \cite{snakeoil_risk_prediction}. This framing and misallocation of responsibility poses a threat to legal due process and social agency \cite{aglorithms_dueprocess}.

Critically, the nature of AI allows these risks to proliferate at an unprecedented scale. For instance, the coding question-and-answer site Stack Overflow banned users from posting answers and code examples generated by ChatGPT due to the vast frequency of generated responses that appear convincing yet are subtly incorrect \cite{chatgpt_stackoverflow}. Similarly, researchers cite increased concern over the risk of AI fueled mis- and dis-information campaigns with vast political and social impacts \cite{misinformation_llm}. Of particular concern is the risk of inextricably embedding different types of social bias in our technological infrastructure \cite{algorithmsofopression,automating_inequality}.

\subsection{Artificially Limited Research Landscapes}
Though perhaps of less existential importance, AI hype also carries risks to the R\&D ecosystem itself. Many of the societal risks described in the previous section are direct consequences of R\&D issues. A failure to interrogate these issues precipitates the risk of permanently entrenching these behaviors into the fabric of AI research.

Dishonest rhetoric around the capabilities of AI systems can lead researchers to believe that we have solved certain problems that we in fact have not; as discussed in the previous section, this can lead to drastic societal consequences. However, it can also profoundly affect the types of research questions that are asked and what areas of development are funded. Wang et al surveyed 387 academic, industry, non-profit, and governmental publications and found widespread critical flaws in predictive optimization systems that use ML to make predictions and decisions about future individual outcomes \cite{wang_against}. Despite research demonstrating that many life outcomes remain inherently unpredictable with currently available data and computing techniques \cite{salganik2020measuring} many of the predictive systems studied by Wang et al were publicly deployed or commercially available. This sets a dangerous precedent that incentivizes more developers to produce similar systems and discourages research into their limitations and shortcomings. It also fosters a culture where younger researchers and students are more likely to erroneously believe that this type of predictive system is inherently trustworthy, further entrenching the idea that alternatives or methods of improvement are not fruitful areas of research. 

This belief in the functionality of outcome predictive AI impacts research questions in individual subfields of AI as well, even some areas typically considered to be under the umbrella of `AI Ethics'. For example, research in AI fairness often presupposes a trade-off between model accuracy and some notion of fairness \cite{balancing_fairness1,fariness_balancing2} and focuses on ways of ensuring AI systems respect existing equal protection laws such as the United States' `4\textbackslash 5ths' rule for treatment of protected classes in hiring decisions \cite{alg_hiring1,alg_hiring2}. However, as Raji et al demonstrates \cite{fallacy}, the utility of this research relies on the assumption that the system in question is accurate and reliable, which, as discussed extensively in previous sections, hyped AI systems often are not. Research on interpretable and explainable AI requires similar assumptions about the fidelity of the system being explained. Furthermore, the current belief in the AI R\&D ecosystem that large-scale models provide the best accuracy and performance \cite{Bengio2007ScalingLA} pushes researchers to focus on black-box explainability techniques to the detriment of other approaches such as developing improved directly interpretable models \cite{rudin_stopblackbox}.

Even if it is not presumed that we have fully solved certain problems, rhetoric that we are very close to solving particular problems, like AGI, with certain approaches, like increasingly larger models and data sets (collectively referred to as `scale', or the `Scaling Hypothesis' \cite{ScalingHypothess}), also profoundly impacts the type of questions researchers are incentivized to pursue. While large scale models have certainly provided state-of-the-art performance on certain tasks, notably language modeling \cite{scaling_gpt1,gpt3_official} and computer vision \cite{cv_scale1,cv_scale2}, and demonstrated impressive abilities, it has not been proven that scale is universally the correct approach to developing more generalizable and performant AI models. In fact, some research has shown that certain substantially smaller models can match large model performance \cite{lottery_tickets} and that the benefits of scale vary based on model architecture and learning task \cite{inductivebias_scale}. Furthermore, extremely large data sets are typically uncurated by necessity, and are therefore unlikely to lead to AI models that can represent diverse view points or truly generalize to real-world tasks \cite{stochastic_parrots}. 

Nonetheless, the field of AI R\&D as a whole is largely focused on scale approaches as evidenced by the coining of the term `foundation models' to position these large scale models as the central and most important mode of AI \cite{foundation_models}. Recently, vast amounts of funding has been poured into research centers and start-up companies focused exclusively on scale approaches. This focus is further driven in part due to large models' performance on benchmark data sets and tasks, which as discussed in Section 3.1, are often not meaningfully connected to real-world operational and societal goals \cite{everything_benchmark}. The confluence of these benchmarks, financial incentives, and hype-fueling research claims creates a feedback loop that serves to trap AI R\&D in a specific and limited focus. 

Rhetoric describing the proximity of scale-driven AGI also impacts what new areas of research are prioritized. Notably, AI Saftey and Alignment, the fields of research concerned with ensuring AGI systems do not pose an existential threat to humanity, have become increasingly popular in the past few years and companies in this space have attracted massive amounts of funding. This stands in stark contrast to the relatively small amount of funding being allocated to research organizations and non-profits focused on alleviating the harms that current AI systems are perpetuating. 

In all areas of research, there is a limit to the available funding and person-power; prioritizing certain research questions necessarily de-prioritizes others, and this selection can have long-lasting and far-reaching consequences. As a field, we must critically consider what questions are worth focusing on in terms of both their potential scientific and societal impacts and what economic and sociological incentives influence our choices of research topics. 

\section{Methods of Addressing and Reducing Hype}
In order to appropriately manage expectations around AI, ensure a robust, responsible, and diverse R\&D ecosystem, and limit the deployment of unsafe and unreliable AI systems, it is critical to change how AI is discussed and drastically reduce the prevalence of AI hype. As described in the previous sections, many factors interact to cultivate a hype-driven feedback loop that impacts how everyone from researchers and developers to the general public perceives and discusses AI. Consequently, combating AI hype requires a multi-faceted approach involving industry, government, academia, and civil society.

\subsection{Diversified Funding and Research}
As discussed in Section 4.2, AI hype serves to focus R\&D on a limited number of questions, typically those that are priorities for technology companies and investors. One way to address this is by encouraging the diversification of AI funding and research at several levels. Although governmental agencies worldwide are investing in AI R\&D, prioritizing that funding towards certain research areas that complement rather than compete with industry will likely be a more effective strategy. There is substantial opportunity for agencies and other funding mechanisms such as research institutes and foundations to support work that re-frames AI R\&D as a fundamentally scientific endeavor. They should concentrate R\&D funding on scientific validity topics like uncertainty and predictive outcome reliability, more rigorous characterization and understanding of system behavior, and non-scale approaches to generalizable AI. Similarly, they should craft funding solicitations that incentivize meaningful interdisciplinary collaboration. By supporting academic research in these areas, this funding also allows students and younger researchers to develop interests in a wider range of topics that will have outward ripple effects in future generations of R\&D.

As members of the AI R\&D ecosystem, we must also take a critical eye towards our field and consider structural changes that can encourage more responsible AI development and more accurate representation of our work. We should make a concerted effort to use more precise language to describe ML and AI; although not all of the underlying mechanisms are precisely understood, rather than resorting to vague or magical language to describe AI systems, we should acknowledge what they are: as Campolo and Crawford put it ``mathematical optimization at scales beyond expert human play'' \cite{enchanted_determinism}. Similarly, we should require a heavier internal burden of proof of the functionality of AI systems before they are deployed. There are multiple proposed frameworks for researchers and developers to interrogate the functionality of their systems \cite{wang_against,leakage} and these can be scaffolded by a wider adoption of core scientific practices like uncertainty estimation and significance testing. Additionally, while emergent behavior in large models may be possible, we should refrain from making unsubstantiated claims about an AI system's reasoning abilities and instead work towards developing rigorous tests to characterize such behavior. As the people initially making claims about an AI system's performance, this burden of proof rests squarely on the researchers and developers \cite{Lundberg2021}. 

Additionally, we should move away from using performance on contrived benchmark data sets as the main metric for success in AI R\&D, as this impedes hypothesis-based research \cite{Hooker1995}, encourages a narrow concentration on short-term objectives \cite{uminsky2020}, and increases the likelihood of reporting overblown performance claims. Raji et al propose several evaluation methods to serve as an addition or alternative to traditional benchmarking. They also argue that insofar as benchmark data sets are useful for characterizing model performance, we must ensure that they are carefully scoped and context-specific \cite{everything_benchmark}. We can also increase the utility and applicability of benchmarks by supporting the development of more curated, collaborative, and sociocultural informed archival data sets \cite{archival_datasets}. 

We must also leave aside technological hubris and embrace meaningful interdisciplinary collaboration that places other domains of knowledge on a level playing field with our own expertise. This will allow us to be more proactive about understanding the societal context and stakeholder value of our work and reduce the likelihood of avoidable feature and label legitimacy issues. Furthermore, it will perhaps discourage some forms of technosolutionism and allow us to consider more diverse imagined futures. 

\subsection{Regulatory Action}
Changes to the R\&D ecosystem are slow, are unlikely to be universally adopted, and are not alone sufficient to tackle the issue of AI hype. While governance and legislation of AI systems is a complex and rapidly evolving field of work, here we provide a few suggestions for external regulation that can reduce the prevalence of unsupported performance claims and increase the reliability of AI discourse. 

Many scholars have suggested standardized third-party audits or testing procedures for evaluating AI system performance, particularly in high-stakes application areas \cite{fda_algorithms,human_oversight_flaws}. If, and only if, these audits do not presuppose system functionality or accept performance claims offered by the developers as a given \cite{fallacy}, requiring them could significantly diminish AI hype by providing a reliable check on technology companies' assertions about their products and discouraging them from overstating AI systems' capabilities in the first place. 

Mandated transparency is also a useful regulatory tool. While requiring transparency into the training data and model development and demanding justification of the scientific basis of technical capability claims enables external audits and allows stakeholders to draw informed conclusions about an AI system regardless of claims put forward by the developers. Although transparency is a complex topic, there is a rich literature around transparent documentation practices to draw from when developing regulatory frameworks for technology \cite{model_cards,nlp_model_cards,interactive_model_cards,datasheets,leakage}. While it is unlikely that regulators could mandate the use of such tools for all AI systems, requiring them for public sector AI systems would help alleviate some of the most dangerous instances of AI hype and perhaps encourage their wider adoption. 

To accelerate the regulatory process, regulators can also rely on existing consumer protection, product liability, and fraud law to address harmful AI hype \cite{fallacy}. Exercising these existing avenues of redress would not only help avoid the dissemination of exaggerated performance and reliability claims, but would also likely impede the willingness of developers to release under-tested AI systems in the first place.

\subsection{Technical Literacy and Best Practices}
Finally, as the consumers and propagators of much of the hype around AI, the media and the general public have a critical role to play in addressing these concerns. Society should not have to rely solely on the developers of a technology to describe its utility \cite{davis2020} and thus it is essential that reporters and writers covering AI have the technical knowledge to avoid reproducing and spreading AI hype. Several excellent frameworks have been proposed to help avoid common pitfalls in AI coverage \cite{journalism_pitfalls,journalism_pitfalls2,journalism_pitfalls3} and it is essential that journalists adopt these techniques to ensure a trustworthy media ecosystem. 

Perhaps the most useful tool for combating AI hype is improved technical literacy amongst all facets of the population, as this will allow everyone to participate in interrogating technological promises and holding AI systems and their developers accountable. Developing a broadly shared vocabulary and understanding of AI is necessary for the appropriate socialization of the technology and enabling and effective distribution of responsibilities and enactment of regulation \cite{Slota2020}. Several public facing initiatives like Better Images of AI and AI Myths have made important steps towards improved public understanding and more meaningful conversation around AI. These efforts must be supported by increased public outreach, opportunities for public engagement in the AI design process, and improved technical K-12 education. Empowering all members of society to meaningfully engage with the AI systems affecting their lives is an essential component of responsible and truthful AI R\&D. This also fosters a more participatory system where researchers can be informed by the actual technological needs and desires of society.

\section{Conclusion}
We have explained how misrepresentations of AI capabilities, or AI hype, originates in many cases from the AI R\&D ecosystem. Hype is further propagated through interactions with the media, funding and regulatory institutions, and the general public, which in turn influence the R\&D ecosystem creating a feedback loop of false narratives, overstated performance claims, misplaced research incentives, and substantial negative societal impact. While no component of the AI hype cycle can be divorced from the others, AI researchers and developers bear a substantial responsibility for recognizing and addressing issues inherent to the field that encourage misleading and unscientific depictions of AI.  

\bibliography{references}

\begin{thebibliography}{115}
\providecommand{\natexlab}[1]{#1}

\bibitem[{Abdalla and Abdalla(2020)}]{bigtech_in_academia}
Abdalla, M.; and Abdalla, M. 2020.
\newblock The Grey Hoodie Project: Big Tobacco, Big Tech, and the threat on academic integrity.
\newblock \emph{CoRR}, abs/2009.13676.

\bibitem[{Abid, Farooqi, and Zou(2021)}]{languagemodel_muslim_bias}
Abid, A.; Farooqi, M.; and Zou, J. 2021.
\newblock Persistent Anti-Muslim Bias in Large Language Models.
\newblock \emph{CoRR}, abs/2101.05783.

\bibitem[{ane Andrew~Lohn, Musser, and Sedova(2021)}]{misinformation_llm}
ane Andrew~Lohn, B.~B.; Musser, M.; and Sedova, K. 2021.
\newblock Truth, Lies, and Automation: How Language Models Could Change Disinformation.
\newblock \emph{Center for Security and Emerging Technology}.

\bibitem[{Antun et~al.(2020)Antun, Renna, Poon, Adcock, and Hansen}]{medicalimages_variability}
Antun, V.; Renna, F.; Poon, C.; Adcock, B.; and Hansen, A.~C. 2020.
\newblock On instabilities of deep learning in image reconstruction and the potential costs of {AI}.
\newblock \emph{Proc. Natl. Acad. Sci. U. S. A.}, 117(48): 30088--30095.

\bibitem[{Atkinson(2016)}]{its_goingto_killus}
Atkinson, R.~D. 2016.
\newblock 'It's Going to Kill Us!' And Other Myths About the Future of Artificial Intelligence.
\newblock \emph{Information Technology \& Innovation Foundation}.

\bibitem[{Bender et~al.(2021)Bender, Gebru, McMillan-Major, and Shmitchell}]{stochastic_parrots}
Bender, E.~M.; Gebru, T.; McMillan-Major, A.; and Shmitchell, S. 2021.
\newblock On the Dangers of Stochastic Parrots: Can Language Models Be Too Big?
\newblock In \emph{Proceedings of the 2021 ACM Conference on Fairness, Accountability, and Transparency}, FAccT '21, 610–623. New York, NY, USA: Association for Computing Machinery.
\newblock ISBN 9781450383097.

\bibitem[{Bender and Koller(2020)}]{nlu_climbing}
Bender, E.~M.; and Koller, A. 2020.
\newblock Climbing towards {NLU}: {On} Meaning, Form, and Understanding in the Age of Data.
\newblock In \emph{Proceedings of the 58th Annual Meeting of the Association for Computational Linguistics}, 5185--5198. Online: Association for Computational Linguistics.

\bibitem[{Bengio and LeCun(2007)}]{Bengio2007ScalingLA}
Bengio, Y.; and LeCun, Y. 2007.
\newblock Scaling learning algorithms towards AI.

\bibitem[{Bickmore et~al.(2018)Bickmore, Trinh, Olafsson, O'Leary, Asadi, Rickles, and Cruz}]{medical_voiceassistant}
Bickmore, T.~W.; Trinh, H.; Olafsson, S.; O'Leary, T.~K.; Asadi, R.; Rickles, N.~M.; and Cruz, R. 2018.
\newblock Patient and Consumer Safety Risks When Using Conversational Assistants for Medical Information: An Observational Study of Siri, Alexa, and Google Assistant.
\newblock \emph{J Med Internet Res}, 20(9): e11510.

\bibitem[{Biddle(2022)}]{chatgpt_bias}
Biddle, S. 2022.
\newblock The Internet’s New Favorite AI Proposes Torturing Iranians and Surveilling Mosques.
\newblock \emph{The Intercept}.

\bibitem[{Bilton(2022)}]{writers_artists_obsolete}
Bilton, N. 2022.
\newblock The New Generation of A.I. Apps Could Make Writers and Artists Obsolete.
\newblock \emph{Vanity Fair}.

\bibitem[{Bommasani et~al.(2021)Bommasani, Hudson, Adeli, Altman, Arora, von Arx, Bernstein, Bohg, Bosselut, Brunskill, Brynjolfsson, Buch, Card, Castellon, Chatterji, Chen, Creel, Davis, Demszky, Donahue, Doumbouya, Durmus, Ermon, Etchemendy, Ethayarajh, Fei{-}Fei, Finn, Gale, Gillespie, Goel, Goodman, Grossman, Guha, Hashimoto, Henderson, Hewitt, Ho, Hong, Hsu, Huang, Icard, Jain, Jurafsky, Kalluri, Karamcheti, Keeling, Khani, Khattab, Koh, Krass, Krishna, Kuditipudi, and et~al.}]{foundation_models}
Bommasani, R.; Hudson, D.~A.; Adeli, E.; Altman, R.~B.; Arora, S.; von Arx, S.; Bernstein, M.~S.; Bohg, J.; Bosselut, A.; Brunskill, E.; Brynjolfsson, E.; Buch, S.; Card, D.; Castellon, R.; Chatterji, N.~S.; Chen, A.~S.; Creel, K.; Davis, J.~Q.; Demszky, D.; Donahue, C.; Doumbouya, M.; Durmus, E.; Ermon, S.; Etchemendy, J.; Ethayarajh, K.; Fei{-}Fei, L.; Finn, C.; Gale, T.; Gillespie, L.; Goel, K.; Goodman, N.~D.; Grossman, S.; Guha, N.; Hashimoto, T.; Henderson, P.; Hewitt, J.; Ho, D.~E.; Hong, J.; Hsu, K.; Huang, J.; Icard, T.; Jain, S.; Jurafsky, D.; Kalluri, P.; Karamcheti, S.; Keeling, G.; Khani, F.; Khattab, O.; Koh, P.~W.; Krass, M.~S.; Krishna, R.; Kuditipudi, R.; and et~al. 2021.
\newblock On the Opportunities and Risks of Foundation Models.
\newblock \emph{CoRR}, abs/2108.07258.

\bibitem[{Branwen(2020)}]{ScalingHypothess}
Branwen, G. 2020.
\newblock The Scaling Hypothesis.

\bibitem[{Bras et~al.(2020)Bras, Swayamdipta, Bhagavatula, Zellers, Peters, Sabharwal, and Choi}]{nlp_adversarial2}
Bras, R.~L.; Swayamdipta, S.; Bhagavatula, C.; Zellers, R.; Peters, M.~E.; Sabharwal, A.; and Choi, Y. 2020.
\newblock Adversarial Filters of Dataset Biases.

\bibitem[{Brown et~al.(2020)Brown, Mann, Ryder, Subbiah, Kaplan, Dhariwal, Neelakantan, Shyam, Sastry, Askell, Agarwal, Herbert-Voss, Krueger, Henighan, Child, Ramesh, Ziegler, Wu, Winter, Hesse, Chen, Sigler, Litwin, Gray, Chess, Clark, Berner, McCandlish, Radford, Sutskever, and Amodei}]{gpt3_official}
Brown, T.~B.; Mann, B.; Ryder, N.; Subbiah, M.; Kaplan, J.; Dhariwal, P.; Neelakantan, A.; Shyam, P.; Sastry, G.; Askell, A.; Agarwal, S.; Herbert-Voss, A.; Krueger, G.; Henighan, T.; Child, R.; Ramesh, A.; Ziegler, D.~M.; Wu, J.; Winter, C.; Hesse, C.; Chen, M.; Sigler, E.; Litwin, M.; Gray, S.; Chess, B.; Clark, J.; Berner, C.; McCandlish, S.; Radford, A.; Sutskever, I.; and Amodei, D. 2020.
\newblock Language Models are Few-Shot Learners.

\bibitem[{Bubeck et~al.(2023)Bubeck, Chandrasekaran, Eldan, Gehrke, Horvitz, Kamar, Lee, Lee, Li, Lundberg, Nori, Palangi, Ribeiro, and Zhang}]{sparks_agi}
Bubeck, S.; Chandrasekaran, V.; Eldan, R.; Gehrke, J.; Horvitz, E.; Kamar, E.; Lee, P.; Lee, Y.~T.; Li, Y.; Lundberg, S.; Nori, H.; Palangi, H.; Ribeiro, M.~T.; and Zhang, Y. 2023.
\newblock Sparks of Artificial General Intelligence: Early experiments with GPT-4.
\newblock arXiv:2303.12712.

\bibitem[{Buolamwini and Gebru(2018)}]{gendershades}
Buolamwini, J.; and Gebru, T. 2018.
\newblock Gender Shades: Intersectional Accuracy Disparities in Commercial Gender Classification.
\newblock In \emph{FAT}.

\bibitem[{Burrell(2016)}]{Burrell2016-gy}
Burrell, J. 2016.
\newblock How the machine `thinks': Understanding opacity in machine learning algorithms.
\newblock \emph{Big Data Soc.}, 3(1): 205395171562251.

\bibitem[{Campolo and Crawford(2020)}]{enchanted_determinism}
Campolo, A.; and Crawford, K. 2020.
\newblock Enchanted Determinism: Power without Responsibility in Artificial Intelligence.
\newblock \emph{Engaging Science, Technology, and Society}, 6: 1.

\bibitem[{Chafkin(2022)}]{selfdriving_nowhere}
Chafkin, M. 2022.
\newblock Even After \$100 Billion, Self-Driving Cars Are Going Nowhere.

\bibitem[{Chiavegatto~Filho, Batista, and Dos~Santos(2021)}]{feature_legitimacy}
Chiavegatto~Filho, A.; Batista, A. F. D.~M.; and Dos~Santos, H.~G. 2021.
\newblock Data leakage in health outcomes prediction with machine learning. Comment on ``prediction of incident hypertension within the next year: Prospective study using statewide electronic health records and machine learning''.
\newblock \emph{J. Med. Internet Res.}, 23(2): e10969.

\bibitem[{Cho, Zala, and Bansal(2022)}]{diffusion_reasoning_bias}
Cho, J.; Zala, A.; and Bansal, M. 2022.
\newblock DALL-Eval: Probing the Reasoning Skills and Social Biases of Text-to-Image Generative Transformers.
\newblock \emph{CoRR}, abs/2202.04053.

\bibitem[{Chouldechova et~al.(2018)Chouldechova, Prado, Fialko, and Vaithianathan}]{afst}
Chouldechova, A.; Prado, D.~B.; Fialko, O.; and Vaithianathan, R. 2018.
\newblock A case study of algorithm-assisted decision making in child maltreatment hotline screening decisions.
\newblock In \emph{FAT}.

\bibitem[{Citron and Pasquale(2014)}]{aglorithms_dueprocess}
Citron, D.~K.; and Pasquale, F. 2014.
\newblock The scored society: Due process for automated predictions.
\newblock \emph{Wash. L. Rev.}, 89: 1.

\bibitem[{CLU(2018)}]{aclu_rekognition}
CLU. 2018.
\newblock ACLU Comment on New Amazon Statement Responding to Face Recognition Technology Test.

\bibitem[{Coyle(2023)}]{writers_strike}
Coyle, J. 2023.
\newblock In Hollywood writers’ battle against AI, humans win (for now).

\bibitem[{Crisan et~al.(2022)Crisan, Drouhard, Vig, and Rajani}]{interactive_model_cards}
Crisan, A.; Drouhard, M.; Vig, J.; and Rajani, N. 2022.
\newblock Interactive Model Cards: A Human-Centered Approach to Model Documentation.
\newblock In \emph{2022 {ACM} Conference on Fairness, Accountability, and Transparency}. {ACM}.

\bibitem[{Davis(2020)}]{davis2020}
Davis, J. 2020.
\newblock How Artifacts Afford: The Power and Politics of Everyday Things.

\bibitem[{Dobbe, Gilbert, and Mintz(2019)}]{normative_uncertainty}
Dobbe, R.; Gilbert, T.~K.; and Mintz, Y. 2019.
\newblock Hard Choices in Artificial Intelligence: Addressing Normative Uncertainty through Sociotechnical Commitments.
\newblock \emph{CoRR}, abs/1911.09005.

\bibitem[{Ebell et~al.(2021)Ebell, Baeza-Yates, Benjamins, Cai, Coeckelbergh, Duarte, Hickok, Jacquet, Kim, Krijger, Macintyre, Madhamshettiwar, Maffeo, Matthews, Medsker, Smith, and Thais}]{aiethics_global_community}
Ebell, C.; Baeza-Yates, R.; Benjamins, R.; Cai, H.; Coeckelbergh, M.; Duarte, T.; Hickok, M.; Jacquet, A.; Kim, A.; Krijger, J.; Macintyre, J.; Madhamshettiwar, P.; Maffeo, L.; Matthews, J.; Medsker, L.; Smith, P.; and Thais, S. 2021.
\newblock Towards intellectual freedom in an AI Ethics Global Community.
\newblock \emph{AI and Ethics}, 1.

\bibitem[{Elish(2019)}]{crumplezones}
Elish, M.~C. 2019.
\newblock Moral crumple zones: Cautionary tales in human-robot interaction.
\newblock \emph{Engaging Science, Technology, and Society}.

\bibitem[{Elish and danah boyd(2018)}]{boyd2018}
Elish, M.~C.; and danah boyd. 2018.
\newblock Situating methods in the magic of Big Data and AI.
\newblock \emph{Communication Monographs}, 85(1): 57--80.

\bibitem[{Eubanks(2018)}]{automating_inequality}
Eubanks, V. 2018.
\newblock \emph{Automating inequality: How high-tech tools profile, police, and punish the poor}.
\newblock St. Martin's Press.

\bibitem[{Fish, Kun, and Lelkes(2016)}]{balancing_fairness1}
Fish, B.; Kun, J.; and Lelkes, A.~D. 2016.
\newblock A Confidence-Based Approach for Balancing Fairness and Accuracy.

\bibitem[{Fleischmann and Wallace(2009)}]{transparency_in_computing}
Fleischmann, K.; and Wallace, W. 2009.
\newblock Ensuring transparency in computational modeling.

\bibitem[{Frankle and Carbin(2018)}]{lottery_tickets}
Frankle, J.; and Carbin, M. 2018.
\newblock The Lottery Ticket Hypothesis: Training Pruned Neural Networks.
\newblock \emph{CoRR}, abs/1803.03635.

\bibitem[{Friedler et~al.(2018)Friedler, Scheidegger, Venkatasubramanian, Choudhary, Hamilton, and Roth}]{fariness_balancing2}
Friedler, S.~A.; Scheidegger, C.; Venkatasubramanian, S.; Choudhary, S.; Hamilton, E.~P.; and Roth, D. 2018.
\newblock A comparative study of fairness-enhancing interventions in machine learning.

\bibitem[{Gao et~al.(2022)Gao, Howard, Markov, Dyer, Ramesh, Luo, and Pearson}]{chatgpt_abstract}
Gao, C.~A.; Howard, F.~M.; Markov, N.~S.; Dyer, E.~C.; Ramesh, S.; Luo, Y.; and Pearson, A.~T. 2022.
\newblock Comparing scientific abstracts generated by ChatGPT to original abstracts using an artificial intelligence output detector, plagiarism detector, and blinded human reviewers.
\newblock \emph{bioRxiv}.

\bibitem[{Gebru et~al.(2018)Gebru, Morgenstern, Vecchione, Vaughan, Wallach, III, and Crawford}]{datasheets}
Gebru, T.; Morgenstern, J.; Vecchione, B.; Vaughan, J.~W.; Wallach, H.~M.; III, H.~D.; and Crawford, K. 2018.
\newblock Datasheets for Datasets.
\newblock \emph{CoRR}, abs/1803.09010.

\bibitem[{Gokhale et~al.(2022)Gokhale, Palangi, Nushi, Vineet, Horvitz, Kamar, Baral, and Yang}]{diffusion_reasoning2}
Gokhale, T.; Palangi, H.; Nushi, B.; Vineet, V.; Horvitz, E.; Kamar, E.; Baral, C.; and Yang, Y. 2022.
\newblock Benchmarking Spatial Relationships in Text-to-Image Generation.

\bibitem[{Grace et~al.(2017)Grace, Salvatier, Dafoe, Zhang, and Evans}]{expected_agi}
Grace, K.; Salvatier, J.; Dafoe, A.; Zhang, B.; and Evans, O. 2017.
\newblock When Will {AI} Exceed Human Performance? Evidence from {AI} Experts.
\newblock \emph{CoRR}, abs/1705.08807.

\bibitem[{Grace et~al.(2024)Grace, Stewart, Sandkühler, Thomas, Weinstein-Raun, and Brauner}]{researcher_survey}
Grace, K.; Stewart, H.; Sandkühler, J.~F.; Thomas, S.; Weinstein-Raun, B.; and Brauner, J. 2024.
\newblock Thousands of AI Authors on the Future of AI.
\newblock arXiv:2401.02843.

\bibitem[{Green(2022)}]{human_oversight_flaws}
Green, B. 2022.
\newblock The flaws of policies requiring human oversight of government algorithms.
\newblock \emph{Computer Law \& Security Review}, 45: 105681.

\bibitem[{Heaven(2022{\natexlab{a}})}]{generativeai_mittech}
Heaven, W.~D. 2022{\natexlab{a}}.
\newblock Generative AI is changing everything. But what’s left when the hype is gone?
\newblock \emph{MIT Technology Review}.

\bibitem[{Heaven(2022{\natexlab{b}})}]{galactica_mit}
Heaven, W.~D. 2022{\natexlab{b}}.
\newblock Why Meta’s latest large language model survived only three days online.
\newblock \emph{MIT Technology Review}.

\bibitem[{Heikkilä(2022)}]{lensa_nude}
Heikkilä, M. 2022.
\newblock The viral AI avatar app Lensa undressed me—without my consent.
\newblock \emph{MIT Technology Review}.

\bibitem[{Hendrycks et~al.(2020)Hendrycks, Burns, Basart, Zou, Mazeika, Song, and Steinhardt}]{mmlu}
Hendrycks, D.; Burns, C.; Basart, S.; Zou, A.; Mazeika, M.; Song, D.; and Steinhardt, J. 2020.
\newblock Measuring Massive Multitask Language Understanding.
\newblock \emph{CoRR}, abs/2009.03300.

\bibitem[{Hooker(1995)}]{Hooker1995}
Hooker, J.~N. 1995.
\newblock Testing heuristics: We have it all wrong.
\newblock \emph{J. Heuristics}, 1(1): 33--42.

\bibitem[{Jack~Brewster(2023)}]{chatgpt_falsehoods}
Jack~Brewster, M.~S., Lorenzo~Arvanitis. 2023.
\newblock The Next Great Misinformation Superspreader: How ChatGPT Could Spread Toxic Misinformation At Unprecedented Scale.
\newblock \emph{NewsGuard}.

\bibitem[{Jo and Gebru(2019)}]{archival_datasets}
Jo, E.~S.; and Gebru, T. 2019.
\newblock Lessons from Archives: Strategies for Collecting Sociocultural Data in Machine Learning.
\newblock \emph{CoRR}, abs/1912.10389.

\bibitem[{Jones, Jones, and Luger(2022)}]{aijournalism_jones}
Jones, B.; Jones, R.; and Luger, E. 2022.
\newblock {AI} `everywhere and nowhere': Addressing the {AI} intelligibility problem in public service journalism.
\newblock \emph{Digit. Journal.}, 10(10): 1731--1755.

\bibitem[{Kaplan et~al.(2020)Kaplan, McCandlish, Henighan, Brown, Chess, Child, Gray, Radford, Wu, and Amodei}]{scaling_gpt1}
Kaplan, J.; McCandlish, S.; Henighan, T.; Brown, T.~B.; Chess, B.; Child, R.; Gray, S.; Radford, A.; Wu, J.; and Amodei, D. 2020.
\newblock Scaling Laws for Neural Language Models.
\newblock \emph{CoRR}, abs/2001.08361.

\bibitem[{Kapoor and Narayanan(2022{\natexlab{a}})}]{snakeoil_risk_prediction}
Kapoor, S.; and Narayanan, A. 2022{\natexlab{a}}.
\newblock The bait and switch behind AI risk prediction tools.
\newblock \emph{AI Snake Oil}.

\bibitem[{Kapoor and Narayanan(2022{\natexlab{b}})}]{leakage}
Kapoor, S.; and Narayanan, A. 2022{\natexlab{b}}.
\newblock Leakage and the Reproducibility Crisis in ML-based Science.

\bibitem[{Kapor and Narayanan(2022)}]{journalism_pitfalls}
Kapor, S.; and Narayanan, A. 2022.
\newblock Eighteen Pitfalls to beware of in AI journalism.

\bibitem[{Knight(2017)}]{alphazero_alien}
Knight, W. 2017.
\newblock Alpha Zero’s “Alien” Chess Shows the Power, and the Peculiarity, of AI.
\newblock \emph{MIT Technology Review}.

\bibitem[{Lecher(2018)}]{welfare_ai}
Lecher, C. 2018.
\newblock What happens when an algorithm cuts your health care.
\newblock \emph{The Verge}.

\bibitem[{Lin and Srikanth(2023)}]{diffusion_reasoning1}
Lin, J.; and Srikanth, M. 2023.
\newblock Diffusion Models as Visual Reasoners.
\newblock In \emph{The AAAI-23 Workshop on Creative AI Across Modalities}.

\bibitem[{Lu et~al.(2017)Lu, Pu, Wang, Hu, and Wang}]{expressivepower_nns2}
Lu, Z.; Pu, H.; Wang, F.; Hu, Z.; and Wang, L. 2017.
\newblock The Expressive Power of Neural Networks: {A} View from the Width.
\newblock \emph{CoRR}, abs/1709.02540.

\bibitem[{Lundberg, Johnson, and Stewart(2021)}]{Lundberg2021}
Lundberg, I.; Johnson, R.; and Stewart, B.~M. 2021.
\newblock What is your estimand? Defining the target quantity connects statistical evidence to theory.
\newblock \emph{Am. Sociol. Rev.}, 86(3): 532--565.

\bibitem[{Marche(2022)}]{chatgpt_college_essay}
Marche, S. 2022.
\newblock The College Essay Is Dead.
\newblock \emph{The Atlantic}.

\bibitem[{McMillan{-}Major et~al.(2021)McMillan{-}Major, Osei, Rodriguez, Ammanamanchi, Gehrmann, and Jernite}]{nlp_model_cards}
McMillan{-}Major, A.; Osei, S.; Rodriguez, J.~D.; Ammanamanchi, P.~S.; Gehrmann, S.; and Jernite, Y. 2021.
\newblock Reusable Templates and Guides For Documenting Datasets and Models for Natural Language Processing and Generation: {A} Case Study of the HuggingFace and {GEM} Data and Model Cards.
\newblock \emph{CoRR}, abs/2108.07374.

\bibitem[{Mitchell, Palmarini, and Moskvichev(2023)}]{conceptarc}
Mitchell, M.; Palmarini, A.~B.; and Moskvichev, A. 2023.
\newblock Comparing Humans, GPT-4, and GPT-4V On Abstraction and Reasoning Tasks.
\newblock arXiv:2311.09247.

\bibitem[{Mitchell et~al.(2018)Mitchell, Wu, Zaldivar, Barnes, Vasserman, Hutchinson, Spitzer, Raji, and Gebru}]{model_cards}
Mitchell, M.; Wu, S.; Zaldivar, A.; Barnes, P.; Vasserman, L.; Hutchinson, B.; Spitzer, E.; Raji, I.~D.; and Gebru, T. 2018.
\newblock Model Cards for Model Reporting.
\newblock \emph{CoRR}, abs/1810.03993.

\bibitem[{Morris et~al.(2024)Morris, Sohl-dickstein, Fiedel, Warkentin, Dafoe, Faust, Farabet, and Legg}]{levels_agi}
Morris, M.~R.; Sohl-dickstein, J.; Fiedel, N.; Warkentin, T.; Dafoe, A.; Faust, A.; Farabet, C.; and Legg, S. 2024.
\newblock Levels of AGI: Operationalizing Progress on the Path to AGI.
\newblock arXiv:2311.02462.

\bibitem[{Nagendran et~al.(2020)Nagendran, Chen, Lovejoy, Gordon, Komorowski, Harvey, Topol, Ioannidis, Collins, and Maruthappu}]{medical_validity_testing}
Nagendran, M.; Chen, Y.; Lovejoy, C.; Gordon, A.; Komorowski, M.; Harvey, H.; Topol, E.; Ioannidis, J.; Collins, G.; and Maruthappu, M. 2020.
\newblock Artificial intelligence versus clinicians: Systematic review of design, reporting standards, and claims of deep learning studies in medical imaging.
\newblock \emph{BMJ}, 368: m689.

\bibitem[{Niven and Kao(2019)}]{nlp_adversarial1}
Niven, T.; and Kao, H.-Y. 2019.
\newblock Probing Neural Network Comprehension of Natural Language Arguments.

\bibitem[{Noble(2018)}]{algorithmsofopression}
Noble, S.~U. 2018.
\newblock Algorithms of oppression.
\newblock In \emph{Algorithms of oppression}. New York University Press.

\bibitem[{Obermeyer et~al.(2019)Obermeyer, Powers, Vogeli, and Mullainathan}]{hospitalbed_bias}
Obermeyer, Z.; Powers, B.; Vogeli, C.; and Mullainathan, S. 2019.
\newblock Dissecting racial bias in an algorithm used to manage the health of populations.
\newblock \emph{Science}, 366(6464): 447--453.

\bibitem[{Oh et~al.(2019)Oh, Dekel, Kim, Mosseri, Freeman, Rubinstein, and Matusik}]{speech2face}
Oh, T.; Dekel, T.; Kim, C.; Mosseri, I.; Freeman, W.~T.; Rubinstein, M.; and Matusik, W. 2019.
\newblock Speech2Face: Learning the Face Behind a Voice.
\newblock \emph{CoRR}, abs/1905.09773.

\bibitem[{Oner et~al.(2020)Oner, Cheng, Lee, and Sung}]{data_segregation}
Oner, M.~U.; Cheng, Y.-C.; Lee, H.~K.; and Sung, W.-K. 2020.
\newblock Training machine learning models on patient level data segregation is crucial in practical clinical applications.

\bibitem[{Prunkl and Whittlestone(2020)}]{beyond_longterm}
Prunkl, C.; and Whittlestone, J. 2020.
\newblock Beyond Near- and Long-Term: Towards a Clearer Account of Research Priorities in AI Ethics and Society.
\newblock Association for Computing Machinery.
\newblock ISBN 9781450371100.

\bibitem[{Raghavan et~al.(2019)Raghavan, Barocas, Kleinberg, and Levy}]{alg_hiring2}
Raghavan, M.; Barocas, S.; Kleinberg, J.~M.; and Levy, K. 2019.
\newblock Mitigating Bias in Algorithmic Employment Screening: Evaluating Claims and Practices.
\newblock \emph{CoRR}, abs/1906.09208.

\bibitem[{Raghu et~al.(2016)Raghu, Poole, Kleinberg, Ganguli, and Sohl-Dickstein}]{expressivepower_nns}
Raghu, M.; Poole, B.; Kleinberg, J.; Ganguli, S.; and Sohl-Dickstein, J. 2016.
\newblock On the Expressive Power of Deep Neural Networks.

\bibitem[{Raji et~al.(2021)Raji, Denton, Bender, Hanna, and Paullada}]{everything_benchmark}
Raji, D.; Denton, E.; Bender, E.~M.; Hanna, A.; and Paullada, A. 2021.
\newblock AI and the Everything in the Whole Wide World Benchmark.
\newblock In Vanschoren, J.; and Yeung, S., eds., \emph{Proceedings of the Neural Information Processing Systems Track on Datasets and Benchmarks}, volume~1.

\bibitem[{Raji and Buolamwini(2019)}]{gendershades_audit}
Raji, I.~D.; and Buolamwini, J. 2019.
\newblock Actionable Auditing: Investigating the Impact of Publicly Naming Biased Performance Results of Commercial AI Products.
\newblock New York, NY, USA: Association for Computing Machinery.
\newblock ISBN 9781450363242.

\bibitem[{Raji et~al.(2022)Raji, Kumar, Horowitz, and Selbst}]{fallacy}
Raji, I.~D.; Kumar, I.~E.; Horowitz, A.; and Selbst, A. 2022.
\newblock The Fallacy of {AI} Functionality.
\newblock In \emph{2022 {ACM} Conference on Fairness, Accountability, and Transparency}. {ACM}.

\bibitem[{Ramesh et~al.(2022)Ramesh, Dhariwal, Nichol, Chu, and Chen}]{cv_scale2}
Ramesh, A.; Dhariwal, P.; Nichol, A.; Chu, C.; and Chen, M. 2022.
\newblock Hierarchical Text-Conditional Image Generation with CLIP Latents.

\bibitem[{Rashida~Richardson(2010)}]{litigating_algorithms}
Rashida~Richardson, V. M.~S., Jason M.~Schultz. 2010.
\newblock Litigating Algorithms 2019 US Report: New Challenges to Government Use of Algorithmic Decision Systems.

\bibitem[{Rayner(2004)}]{novelty_trap}
Rayner, S. 2004.
\newblock The Novelty Trap: Why Does Institutional Learning about New Technologies Seem So Difficult?
\newblock \emph{Industry and Higher Education}, 18: 349--355.

\bibitem[{Rudin(2018)}]{rudin_stopblackbox}
Rudin, C. 2018.
\newblock Stop Explaining Black Box Machine Learning Models for High Stakes Decisions and Use Interpretable Models Instead.

\bibitem[{Salganik et~al.(2020)Salganik, Lundberg, Kindel, Ahearn, Al-Ghoneim, Almaatouq, Altschul, Brand, Carnegie, Compton et~al.}]{salganik2020measuring}
Salganik, M.~J.; Lundberg, I.; Kindel, A.~T.; Ahearn, C.~E.; Al-Ghoneim, K.; Almaatouq, A.; Altschul, D.~M.; Brand, J.~E.; Carnegie, N.~B.; Compton, R.~J.; et~al. 2020.
\newblock Measuring the predictability of life outcomes with a scientific mass collaboration.
\newblock \emph{Proceedings of the National Academy of Sciences}, 117(15): 8398--8403.

\bibitem[{Sambasivan and Veeraraghavan(2022)}]{deskilling}
Sambasivan, N.; and Veeraraghavan, R. 2022.
\newblock The Deskilling of Domain Expertise in AI Development.
\newblock New York, NY, USA: Association for Computing Machinery.
\newblock ISBN 9781450391573.

\bibitem[{Sculley et~al.(2015)Sculley, Holt, Golovin, Davydov, Phillips, Ebner, Chaudhary, Young, Crespo, and Dennison}]{technical_debt}
Sculley, D.; Holt, G.; Golovin, D.; Davydov, E.; Phillips, T.; Ebner, D.; Chaudhary, V.; Young, M.; Crespo, J.-F.; and Dennison, D. 2015.
\newblock Hidden Technical Debt in Machine Learning Systems.
\newblock In Cortes, C.; Lawrence, N.; Lee, D.; Sugiyama, M.; and Garnett, R., eds., \emph{Advances in Neural Information Processing Systems}, volume~28. Curran Associates, Inc.

\bibitem[{Sezgin et~al.(2020)Sezgin, Huang, Ramtekkar, and Lin}]{va_health}
Sezgin, E.; Huang, Y.; Ramtekkar, U.; and Lin, S. 2020.
\newblock Readiness for voice assistants to support healthcare delivery during a health crisis and pandemic.
\newblock \emph{NPJ Digital Medicine}, 3(1): 122.

\bibitem[{Shah and Bender(2022)}]{situating_search}
Shah, C.; and Bender, E.~M. 2022.
\newblock Situating Search.
\newblock CHIIR '22, 221–232. New York, NY, USA: Association for Computing Machinery.
\newblock ISBN 9781450391863.

\bibitem[{Shardlow and Przybyła(2022)}]{llm_sentience}
Shardlow, M.; and Przybyła, P. 2022.
\newblock Deanthropomorphising NLP: Can a Language Model Be Conscious?

\bibitem[{Shneiderman(2022)}]{journalism_pitfalls2}
Shneiderman, B. 2022.
\newblock Guidelines for journalists and editors about reporting on robots, AI, and computers.

\bibitem[{Silver et~al.(2017)Silver, Schrittwieser, Simonyan, Antonoglou, Huang, Guez, Hubert, Baker, Lai, Bolton, Chen, Lillicrap, Hui, Sifre, van~den Driessche, Graepel, and Hassabis}]{alphazero}
Silver, D.; Schrittwieser, J.; Simonyan, K.; Antonoglou, I.; Huang, A.; Guez, A.; Hubert, T.; Baker, L.; Lai, M.; Bolton, A.; Chen, Y.; Lillicrap, T.; Hui, F.; Sifre, L.; van~den Driessche, G.; Graepel, T.; and Hassabis, D. 2017.
\newblock Mastering the game of Go without human knowledge.
\newblock \emph{Nature}, 550(7676): 354--359.

\bibitem[{Simonite(2015)}]{baidu_cheating}
Simonite, T. 2015.
\newblock Why and How Baidu Cheated an Artificial Intelligence Test.
\newblock \emph{MIT Technology Review}.

\bibitem[{Sivadas and Argoub(2021)}]{journalism_pitfalls3}
Sivadas, L.; and Argoub, S. 2021.
\newblock How to report effectively on artificial intelligence.

\bibitem[{Slota et~al.(2020)Slota, Fleischmann, Greenberg, Verma, Cummings, Li, and Shenefiel}]{Slota2020}
Slota, S.~C.; Fleischmann, K.~R.; Greenberg, S.; Verma, N.; Cummings, B.; Li, L.; and Shenefiel, C. 2020.
\newblock Good systems, bad data?: Interpretations of AI hype and failures.
\newblock \emph{Proc. Assoc. Inf. Sci. Technol.}, 57(1).

\bibitem[{Stark and Hutson(2021)}]{ai_physigonomy}
Stark, L.; and Hutson, J. 2021.
\newblock Physiognomic artificial intelligence.
\newblock \emph{SSRN Electron. J.}

\bibitem[{Stilgoe(2018)}]{ml_selfdriving}
Stilgoe, J. 2018.
\newblock Machine learning, social learning and the governance of self-driving cars.
\newblock \emph{Social Studies of Science}, 48(1): 25--56.
\newblock PMID: 29160165.

\bibitem[{Stilgoe(2019)}]{selfdriving_faroff}
Stilgoe, J. 2019.
\newblock Self-driving cars will take a while to get right.
\newblock \emph{Nature Machine Intelligence}, 1.

\bibitem[{Stilgoe and Mladenovi{\'c}(2022)}]{politics_autonomousvehicles}
Stilgoe, J.; and Mladenovi{\'c}, M. 2022.
\newblock The politics of autonomous vehicles.
\newblock \emph{Humanit. Soc. Sci. Commun.}, 9(1).

\bibitem[{Tay et~al.(2022)Tay, Dehghani, Abnar, Chung, Fedus, Rao, Narang, Tran, Yogatama, and Metzler}]{inductivebias_scale}
Tay, Y.; Dehghani, M.; Abnar, S.; Chung, H.~W.; Fedus, W.; Rao, J.; Narang, S.; Tran, V.~Q.; Yogatama, D.; and Metzler, D. 2022.
\newblock Scaling Laws vs Model Architectures: How does Inductive Bias Influence Scaling?

\bibitem[{Taylor et~al.(2022)Taylor, Kardas, Cucurull, Scialom, Hartshorn, Saravia, Poulton, Kerkez, and Stojnic}]{galactica}
Taylor, R.; Kardas, M.; Cucurull, G.; Scialom, T.; Hartshorn, A.; Saravia, E.; Poulton, A.; Kerkez, V.; and Stojnic, R. 2022.
\newblock Galactica: A Large Language Model for Science.

\bibitem[{Thomas and Uminsky(2020)}]{uminsky2020}
Thomas, R.~L.; and Uminsky, D. 2020.
\newblock The Problem with Metrics is a Fundamental Problem for {AI}.
\newblock \emph{CoRR}, abs/2002.08512.

\bibitem[{Tiku(2022)}]{ladma}
Tiku, N. 2022.
\newblock The Google engineer who thinks the company’s AI has come to life.
\newblock \emph{The Washington Post}, 11.

\bibitem[{Tutt(2016)}]{fda_algorithms}
Tutt, A. 2016.
\newblock An {FDA} for algorithms.
\newblock \emph{SSRN Electron. J.}

\bibitem[{UNESCO(2021)}]{unesco_science_report}
UNESCO. 2021.
\newblock UNESCO Science Report: The race against time for smarter development.

\bibitem[{Valmeekam et~al.(2023)Valmeekam, Marquez, Olmo, Sreedharan, and Kambhampati}]{valmeekam2023planbench}
Valmeekam, K.; Marquez, M.; Olmo, A.; Sreedharan, S.; and Kambhampati, S. 2023.
\newblock PlanBench: An Extensible Benchmark for Evaluating Large Language Models on Planning and Reasoning about Change.
\newblock In \emph{Thirty-seventh Conference on Neural Information Processing Systems Datasets and Benchmarks Track}.

\bibitem[{VINCENT(2022)}]{chatgpt_stackoverflow}
VINCENT, J. 2022.
\newblock AI-generated answers temporarily banned on coding Q \&A site Stack Overflow.
\newblock \emph{The Verge}.

\bibitem[{Vinsel(2021)}]{critics_hype}
Vinsel, L. 2021.
\newblock You’re Doing It Wrong: Notes on Criticism and Technology Hype.

\bibitem[{Wan et~al.(2021)Wan, Xia, Lo, and Murphy}]{ml_dev_practices}
Wan, Z.; Xia, X.; Lo, D.; and Murphy, G.~C. 2021.
\newblock How does Machine Learning Change Software Development Practices?
\newblock \emph{IEEE Transactions on Software Engineering}, 47(9): 1857--1871.

\bibitem[{Wang et~al.(2022)Wang, Kapoor, Barocas, and Narayanan}]{wang_against}
Wang, A.; Kapoor, S.; Barocas, S.; and Narayanan, A. 2022.
\newblock Against Predictive Optimization: On the Legitimacy of Decision-Making Algorithms that Optimize Predictive Accuracy.
\newblock \emph{Available at SSRN}.

\bibitem[{West(2011)}]{brookings_tech_economy}
West, D.~M. 2011.
\newblock Technology and the Innovation Economy.
\newblock \emph{Center for Technology Innovation at the Brookings Institute}.

\bibitem[{Wilson et~al.(2021)Wilson, Ghosh, Jiang, Mislove, Baker, Szary, Trindel, and Polli}]{alg_hiring1}
Wilson, C.; Ghosh, A.; Jiang, S.; Mislove, A.; Baker, L.; Szary, J.; Trindel, K.; and Polli, F. 2021.
\newblock Building and Auditing Fair Algorithms: A Case Study in Candidate Screening.
\newblock New York, NY, USA: Association for Computing Machinery.
\newblock ISBN 9781450383097.

\bibitem[{Witteveen and Andrews(2022)}]{diffusion_prompt_engineering}
Witteveen, S.; and Andrews, M. 2022.
\newblock Investigating Prompt Engineering in Diffusion Models.

\bibitem[{Wong et~al.(2021)Wong, Otles, Donnelly, Krumm, McCullough, DeTroyer-Cooley, Pestrue, Phillips, Konye, Penoza, Ghous, and Singh}]{sepsis}
Wong, A.; Otles, E.; Donnelly, J.~P.; Krumm, A.; McCullough, J.; DeTroyer-Cooley, O.; Pestrue, J.; Phillips, M.; Konye, J.; Penoza, C.; Ghous, M.; and Singh, K. 2021.
\newblock External validation of a widely implemented proprietary sepsis prediction model in hospitalized patients.
\newblock \emph{JAMA Intern. Med.}, 181(8): 1065--1070.

\bibitem[{Xu et~al.(2020)Xu, Ju, Li, Boureau, Weston, and Dinan}]{chatbot_safetyrecs}
Xu, J.; Ju, D.; Li, M.; Boureau, Y.; Weston, J.; and Dinan, E. 2020.
\newblock Recipes for Safety in Open-domain Chatbots.
\newblock \emph{CoRR}, abs/2010.07079.

\bibitem[{Yu et~al.(2022)Yu, Xu, Koh, Luong, Baid, Wang, Vasudevan, Ku, Yang, Ayan, Hutchinson, Han, Parekh, Li, Zhang, Baldridge, and Wu}]{cv_scale1}
Yu, J.; Xu, Y.; Koh, J.~Y.; Luong, T.; Baid, G.; Wang, Z.; Vasudevan, V.; Ku, A.; Yang, Y.; Ayan, B.~K.; Hutchinson, B.; Han, W.; Parekh, Z.; Li, X.; Zhang, H.; Baldridge, J.; and Wu, Y. 2022.
\newblock Scaling Autoregressive Models for Content-Rich Text-to-Image Generation.

\bibitem[{Zech et~al.(2018)Zech, Badgeley, Liu, Costa, Titano, and Oermann}]{medicalimages_hospitaltransfer}
Zech, J.~R.; Badgeley, M.~A.; Liu, M.; Costa, A.~B.; Titano, J.~J.; and Oermann, E.~K. 2018.
\newblock Variable generalization performance of a deep learning model to detect pneumonia in chest radiographs: A cross-sectional study.
\newblock \emph{PLoS Med.}, 15(11): e1002683.

\bibitem[{Zhang et~al.(2023)Zhang, Maslej, Brynjolfsson, Etchemendy, Lyons, Manyika, Ngo, Niebles, Sellitto, Sakhaee, Shoham, Clark, , and Perrault}]{hai_report}
Zhang, D.; Maslej, N.; Brynjolfsson, E.; Etchemendy, J.; Lyons, T.; Manyika, J.; Ngo, H.; Niebles, J.~C.; Sellitto, M.; Sakhaee, E.; Shoham, Y.; Clark, J.; ; and Perrault, R. 2023.
\newblock The AI Index 2023 Annual Report.
\newblock \emph{Stanford Institute for Human-Centered AI}.

\end{thebibliography}

\end{document}